\documentclass{ws-procs9x6}
\pdfoutput=1
\def\lsim{\mathrel{\rlap{\lower4pt\hbox{\hskip1pt$\sim$}}
    \raise1pt\hbox{$<$}}}         
\def\gsim{\mathrel{\rlap{\lower4pt\hbox{\hskip1pt$\sim$}}
    \raise1pt\hbox{$>$}}}         
\begin{document}
\begin{flushright}
INT PUB 08-30
\end{flushright}

\title{EDWARD TELLER AND NUCLEI:\\
ALONG THE TRAIL TO THE NEUTRINO}

\author{W. C. HAXTON$^*$}

\address{Institute for Nuclear Theory and Department of Physics, University of Washington,\\
Box 351550, Seattle, WA 98195\\
$^*$E-mail: haxton@phys.washington.edu\\
www.int.washington.edu/haxton.html}

\begin{abstract}
I discuss two of Edward Teller's contributions to nuclear physics, the introduction
of the Gamow-Teller operator in $\beta$ decay and the formulation of the 
Goldhaber-Teller model for electric dipole transitions, in the context of efforts
to understand the weak interaction and the nature of the neutrino.
\end{abstract}


\bodymatter

\section{Introduction}
\label{sec:sec1}
It is a great pleasure to take part in the Edward Teller Centennial Symposium and
to have this opportunity to describe Dr. Teller's contributions to electroweak nuclear physics.
His career in physics began just at the time that the nature of $\beta$ decay and
the likely existence of the neutrino were first becoming clear.  Edward's own
contributions to this story are very significant and involve both neutrino properties
and the role of weak interactions in astrophysics.

Experiments on radioactive nuclei had, by the late-1920s, demonstrated that
the positrons emitted in $\beta$ decay were produced in a continuous spectrum,
carrying off on average only about half of the nuclear
decay energy \cite{brown}.  James Chadwick first obtained this result in 1914 from studies of the beta
decay of $^{214}$Pb.  A particularly definitive calorimetry measurement was done by Ellis and
Wooster in 1927 and confirmed and improved by Meitner and Orthman in 1930.   
Speculative explanations included Niels Bohr's
suggestion that Einstein's mass/energy equivalence might not hold in the
``new quantum mechanics," and Chadwick's observation that perhaps some
unobserved radiation accompanied the positron.

In 1930 Wolfgang Pauli, who Bohr had once described as a ``genius, comparable
perhaps only to Einstein himself," hypothesized that an unobserved, neutral, spin-1/2
``neutron," later renamed the neutrino by Fermi,
accounted for the apparent anomaly -- a new particle with a mass less than
1\% that of the proton.  His suggestion came in a letter to the participants
of a conference in Tuebingen that began with ``Liebe Radioaktive Damen und Herren!"
Reflecting perhaps the more conservative nature of theorists of that era, he later 
worried: ``I have done a terrible thing.  I have postulated a particle that cannot be detected."
Pauli first public presentation on the neutrino did not come until the 1933 7th Solvay
Conference \cite{pauli}.

\begin{figure}
\begin{center}
\includegraphics[width=11cm]{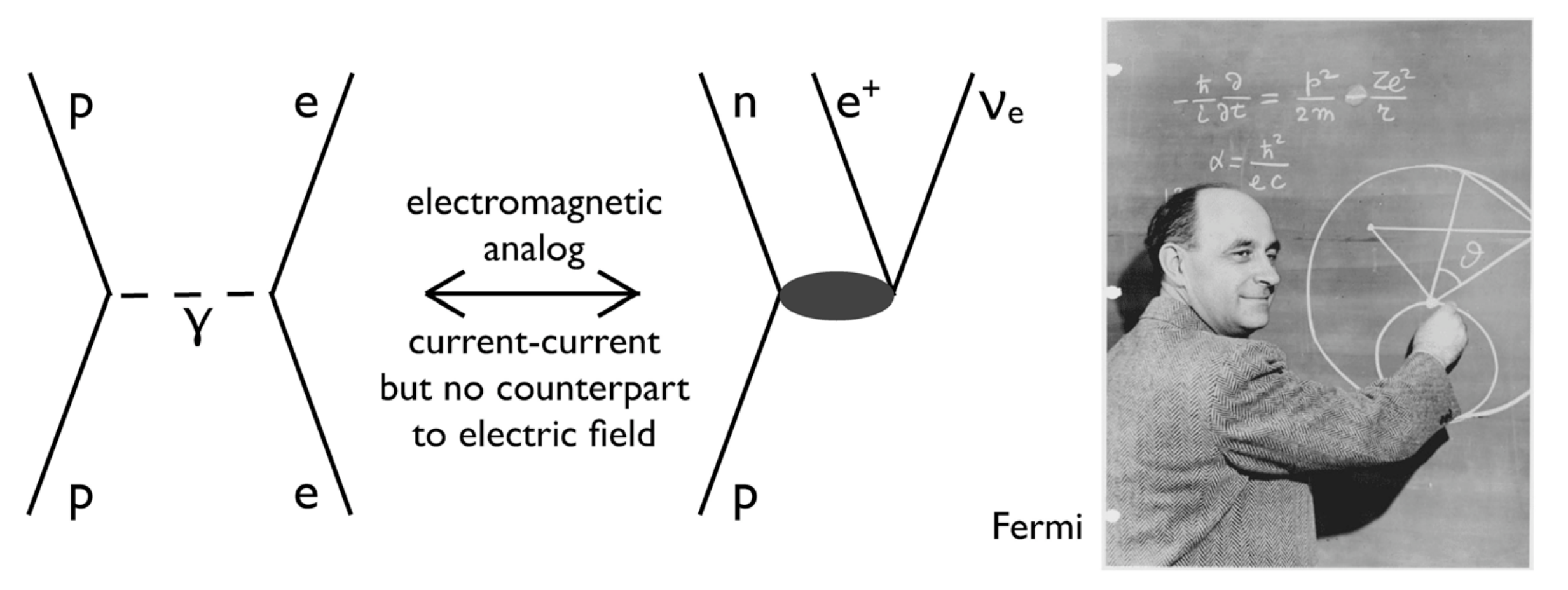}
\end{center}
\caption{ The four-fermion interaction Fermi proposed as a model for $\beta$ decay,
an analog of the electromagnetic interaction apart from the absence of the 
electromagnetic field. Photo of Fermi (note the blackboard error) courtesy of Argonne
National Laboratory.}
\label{fig:Fermi}
\end{figure}

Today's neutron was found by Chadwick in 1932.  Fermi, who had followed Pauli's
suggestion and other developments in $\beta$ decay, in 1934 proposed a model of the
weak interaction, based on an analogy with electromagnetism \cite{fermi}.  As shown
in Fig.~\ref{fig:Fermi}, a proton bound in a nucleus is transformed into a neutron, with the emission of a positron
and $\nu_e$.   The interaction occurs at a point, so Fermi's model has no counterpart
to the electromagnetic field.  Remarkably, apart from axial currents and the
associated parity violation, this description
is the correct low-energy limit of today's standard model.  It anticipated future
developments in which the electromagnetic and weak interactions would be unified,
in the standard model and in the nuclear response to these interactions. 

\section{Teller \cite{teller} and Gamow \cite{gamow}}
Edward Teller completed his PhD in Leipzig in the same year, 1930, that Pauli
made his suggestion of the neutrino.   Teller  worked with Werner Heisenberg on
the quantum mechanics of molecular hydrogen, and had benefited from the many
visitors attracted to the university.  One of special note was George Gamow, who visited
in 1930, accompanied by Lev Landau.  Gamow had described barrier penetration
in 1928 to explain $\alpha$ decay, setting the stage for later treatments of nuclear
astrophysics and the Gamow peak.

Teller became an assistant at G\"{o}ttingen in 1931, helping Eucken and Franck.  Through
his friendship with Czech physicist George Placzek, Teller received an invitation to
visit Fermi in the summer of 1932.  Teller describes the circumstances \cite{archive}:

\begin{quote}
{\it Placzek wanted to continue with his work in the holidays while G\"{o}ttingen was closed.
I wanted to go home to Budapest and he said:  ``No. I, Placzek, want to visit with Fermi in Rome
and you come along."  In a way, I was interested but I had just started to make my own money,
needed little help from home.  Didn't know quite how to pay for my stay in Rome.  ``Oh said
Placzek -- I will take care of that.  I'll ask Fermi."  Here I get a copy of a letter from Fermi, that
he has written to appropriate authorities in Hungary -- I hear that Dr. Teller is considering to visit
Rome for a few weeks.  He is a very famous physicist, and I want his cooperation.
Could you please help him to get to Rome and to stay there?  Fermi and I had never met.
That he had reason to consider me as a famous physicist was, to say the very least,
an impudent exaggeration.  But what makes the story particularly enjoyable for me 
is that together with the copy of the letter to the Hungarian authorities, I got, attached, a little
note from Fermi:  Dear Dr. Teller, I am sending you this copy.  I want you to know that actually
I would be really very happy to see you in Rome.  So he took back the exaggeration but 
replaced it with a, a premature offer to friendship which, of course, became a very real friendship
in the course of time.}
\end{quote}

This must have been an exciting summer to have visited Fermi.  A third visitor, Hans Bethe,
had also been invited.  The neutron had been discovered a few months earlier.  (Teller later
notes that Fermi's studies of neutron interactions with nuclei, begun in 1934 and revealing
the odd activation of uranium, could have profoundly affected world politics, had Fermi
correctly interpreted the results.)

Teller's friend Gamow had resolved to leave Russia by this point.  In 1932 he made 
two attempts to defect by kayak with his wife, Lyubov
Vokhminzeva, first via the Black Sea to Turkey, and later from Murmansk to
Norway.  Both attempts failed because of poor weather conditions.  But in 1933 a simpler
opportunity arose when both were granted visas
to leave Leningrad to attend the 7th Solvay
Conference -- they same conference where Pauli finally discussed the neutrino.  By
1934 Gamow had moved to the US to join the faculty at George
Washington University.

Teller's path led him, after two years in G\"{o}ttingen, to England in 1933, a move facilitated
by the Jewish Rescue Committee.  He soon arrived in Copenhagen to spend a year with
Niels Bohr.  In February 1934 he married Augusta Maria Harkanyi, the sister of a friend.
In 1935, at the behest of Gamow, Teller was invited to become a professor at George Washington
University.  He remained there for the next six years.  In March 1936 Gamow and Teller 
submitted a paper on $\beta$ decay to the Physical Review arguing that a spin-dependent
interaction -- an axial current -- was required to account for observed selection rules.

\section{Gamow-Teller Transitions}
The Gamow-Teller paper's abstract reads \cite{GTpaper}
\begin{quote}
{\it \S1.  The selection rules for $\beta$-transformations are stated on the basis of the
neutrino theory outlined by Fermi.  If it is assumed that the spins of the heavy particles
have a direct effect on the disintegration these rules are modified.  \S2. It is shown that
whereas the original selection rules of Fermi lead to difficulties if one tries to assign
spins to the members of the thorium family the modified selection rules are in agreement
with the available experimental evidence.}
\end{quote}
Fermi's treatment of $\beta$ decay in analogy with electromagnetism predicts that
weak transitions will obey the selection rules of a vector current, which in contemporary
notation are displayed in Table~\ref{table:one}.  In the nonrelativistic, long-wavelength
limit -- the allowed limit -- only the $\mu=0$ charge operator arises, with selection rules
\[ \Delta J=0~~~~\Delta \pi=0,~~\mathrm{e.g.,}~~ 0^+ \leftrightarrow 0^+ .\]
The Gamow-Teller paper also notes that Fermi had introduced, to ensure relativistic
invariance, the velocity operator $\vec{p}/M$, with selection rules
\[ \Delta J=0,\pm 1 \mathrm{~but~no~0\leftrightarrow 0}~~~~\Delta \pi=1,~~\mathrm{e.g.,}~~ 0^+ \leftrightarrow 1^- .\]
These operators are shown in the first row of Table~\ref{table:one}.

\begin{table}
\tbl{Long-wavelength vector-current operators, first-order corrections proportional
to $k \vec{r}$, and long-wavelength axial-vector operators. }
{\begin{tabular}{c|l|c|c|c}
\cline{2-4}
 ~~~~~~~~~~~~~~~&& $\mu$=0 & $\mu$=1,2,3&~~~~~~~~~~~~~~~ \\
 \cline{2-4}
&$J_\mu^V(\vec{r})$ & 1 & $\vec{p}/M$ &\\
&$J_\mu^V(\vec{r}) ~e^{i \vec{k} \cdot \vec{r}}$ & $k \vec{r}$ & $k [\vec{r} \otimes \vec{p}/M]_{J=0,1,2}$ &\\
&$J_\mu^A(\vec{r})$ & $g_A \vec{\sigma} \cdot \vec{p}/M$ & $g_A \vec{\sigma} $&\\
\cline{2-4}
\end{tabular}}
\label{table:one}
\end{table}

Gamow and Teller considered an additional operator arising from the three-momentum
transfered to the nucleus, obtained from a spherical harmonic expansion of $e^{i \vec{k} \cdot
\vec{r}}$.   The first entry
in the second row of Table \ref{table:one}, the dipole operator, is
first-forbidden, suppressed by one power of $\vec{k}$, with
the same selection rules as the velocity operator.  Its matrix elements
can be related to those of the velocity
operator through current conservation.  (This operator is the subject of another
famous Teller paper, discussed later in this talk.)  Gamow and Teller argued that $\beta$-decay rates
for such an operator would be suppressed typically by $\sim 10^{-3}$.   They did not
discuss explicitly the second-forbidden operator appearing in the second row,
which could mediate $0^+ \leftrightarrow 1^+$ transition, but would be suppressed by an
additional factor of $\sim 10^{-2}$ due to relativity, as nucleon velocities in the nucleus
are of order $|\vec{v}/c| \sim 0.1$.  But certainly they would have recognized that this
operator exists and is numerically insignificant.

Because the selection rules for allowed and first-forbidden operators built on the vector current
could not explain strong magnetic transitions seen in $\beta$-decay experiments, Gamow and Teller
introduced second allowed operator, the spin operator corresponding to the space-like part
of an axial-vector current $J_\mu^A(\vec{r})$, as shown in the third row of Table~\ref{table:one}. The selection rules for this operator are
\[ \Delta J=0,\pm 1 \mathrm{~but~no~0\leftrightarrow 0}~~~~\Delta \pi=0,~~\mathrm{e.g.,}~~ 0^+ \leftrightarrow 1^+ .\]

Gamow and Teller designated the Fermi operator -- the vector charge operator -- $M1$ and
the new Gamow-Teller operator - the axial three-current -- $M2$, noting
\begin{quote}
{\it Either the matrix element M1 or the matrix element M2 or finally a linear combination
of M1 and M2 will have to be used to calculate the probabilities of the $\beta$-disintegrations.
If the third possibility is the correct one, and the two coefficients in the linear combination
have the same order of magnitude, then all transitions [satisfying the selection rules] would
now lie on the first Sargent curve.}
\end{quote}
The first Sargent curve is the term then used to designate a strong, allowed
transition.

This observation is remarkable in several ways.  It gives the correct allowed rate in the absence
of polarization
\[ \omega \sim |\langle f | 1 | i \rangle |^2 + g_A^2 |\langle f | \vec{\sigma} | i \rangle|^2. \]
Gamow and Teller suggested that the vector and axial coupling might be similar
in magnitude.  Today's standard model describes the coupling to the underlying quarks 
as $V-A$.   Also, while the following issue does not arise for rates -- the focus
of the Gamow-Teller paper --  one could accomplish
the suggested generalization of Fermi's theory in two ways.  In a modern notation,
\[
 J_\mu^V J^{V \mu} \rightarrow \left\{ \begin{array}{l} J_\mu^V J^{V \mu}+J_\mu^A J^{A \mu} \\
  (J_\mu^V - J_\mu^A)(J^{V \mu} - J^{A \mu}) \end{array} \right. \]
That is, one could add an  axial current-current interaction to Fermi's vector-vector interaction,
 or alternatively generalize the current to $V-A$.
The latter would have anticipated by 20 years parity violation in the weak interaction.
While Gamow and Teller do not comment explicitly on this issue or mention parity
violation,  their description of the matrix elements $M1$ and $M2$ is curious.  They 
speak of a linear combination of $M1$ and $M2$ -- that is, a sum of vector and axial-vector
amplitudes, which they suggest would be of comparable importance.  They were
remarkably close to the standard model.

The introduction of the Gamow-Teller operator was critical to efforts in the 1930s to
understand the mechanisms for hydrogen burning in main-sequence stars.  The initiating
step in the pp chain synthesis of $^4$He
\[ p + p~ (L=0~S=0~T=1) \rightarrow d~ (L=0 ~S=1 ~T=0) + e^+ + \nu_e \]
is a Gamow-Teller transition.

The structure of the weak interaction -- the amplitudes that might be constructed from
vector (V), axial-vector (A), scalar (S), pseudoscalar (P), and tensor (T) terms at low energies --
was not fully resolved until 30 years later.   Teller had another connection with this story,
through his University of Chicago student C. N. Yang (1946-9).   In 1957 Lee and Yang
pointed out that parity conservation was poorly established in the weak interaction,
and that its violation might explain puzzling decay properties of neutral kaons.   Quickly
following this suggestion Wu, Ambler, Hayward, Hoppes, and Hudson
found an angular asymmetry in the $\beta$s produced
in the decay of polarized $^{60}$Co, and Garwin, Lederman, and Weinrich found highly
polarized  muons in pion $\beta$ decay.   Finally, Goldhaber,
Grodzins, and Sunyar demonstrated that $\beta$ decay neutrinos are left-handed,
ruling out possibilities like S+T in favor of V-A.

\section{Particles, Antiparticles, and Neutrino Mass}
While the $V-A$ nature of the weak interaction may seem an old story, it has a modern
connection to some very important open questions regarding the neutrino.  The neutrino
is unique among standard model fermions in that it lacks a charge or any other
additively conserved quantum number.    Such quantum numbers reverse sign under
particle-antiparticle conjugation, and thus distinguish particles from their antiparticles.
Thus we know that the electron has a distinct antiparticle, the position with its opposite charge.
But in the case of the neutrino, the need for a distinct antiparticle is unclear:  
it is possible that the neutrino is its own antiparticle.

\begin{figure}
\begin{center}
\includegraphics[width=11cm]{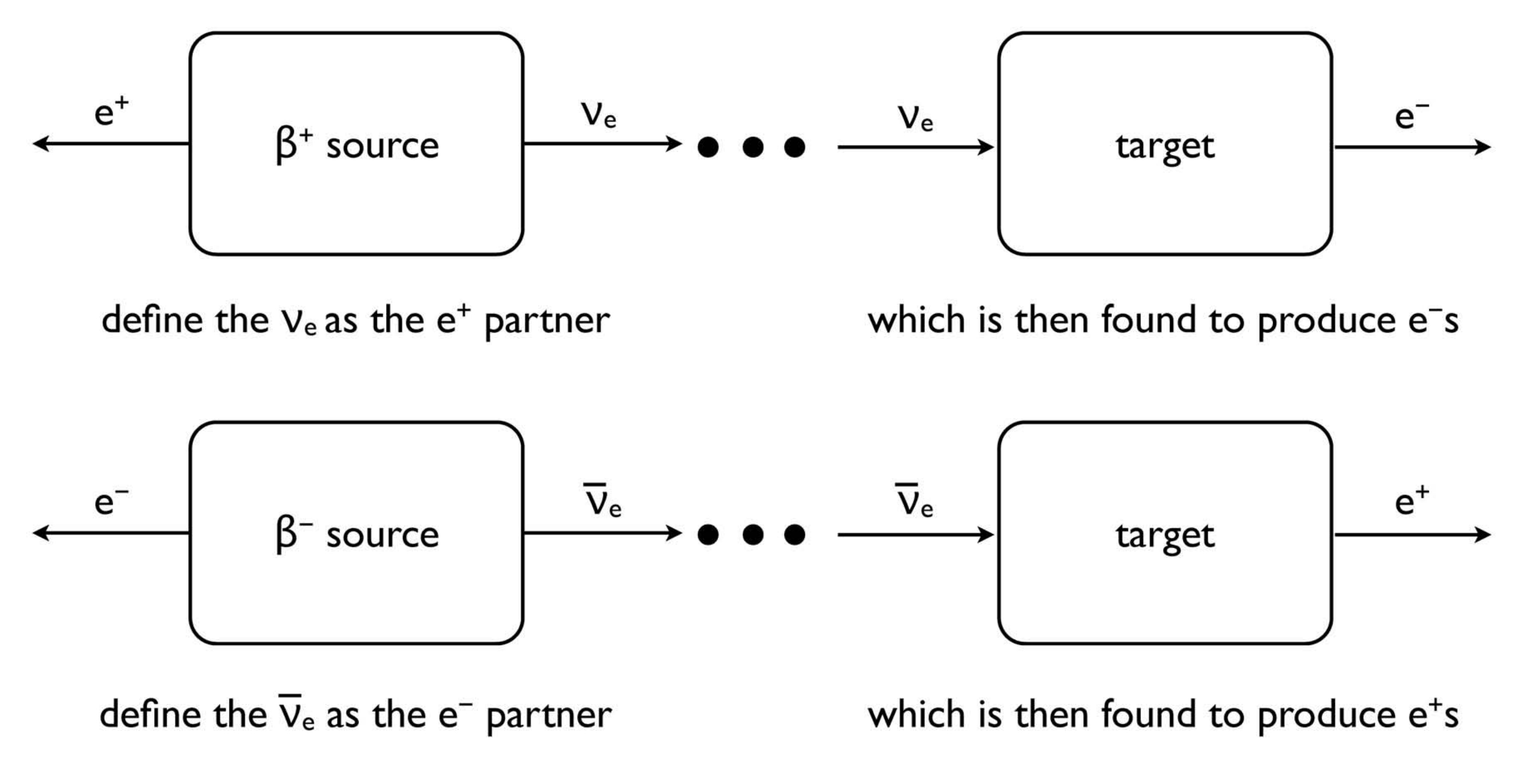}
\end{center}
\caption{ An operational test of the identity of the $\nu_e$ and $\bar{\nu}_e$, prior to
the discovery of parity nonconservation.  If one were to define the $\nu_e$ as the partner
of the positron in $\beta^+$ decay and the $\bar{\nu}_e$ as the partner of the electron
in $\beta^-$ decay, then these neutrinos would appear to be distinct operationally,
when their interactions in targets are tested.  This would seem to require the introduction of
a quantum number like lepton number to distinguish the $\nu$ from the $\bar{\nu}$.}
\label{fig:nue}
\end{figure}

This might prompt one to do the gedanken experiment illustrated in 
the top panel of Fig.~\ref{fig:nue}.  The first step
is to define the $\nu_e$ as the particle that accompanies the positron produced by a
$\beta^+$ source.  The second step is to determine what that particle does when it
strikes a target.  One finds it produces $e^-$s.

Next define the $\bar{\nu}_e$ as the particle that accompanies the electron
produced by a $\beta^-$ source.  Then one finds that this particle, on  striking a
target, produces an $e^+$.  From a comparison of the two experiments it appears
that the $\nu_e$ and $\bar{\nu}_e$ are distinct particles: they produce different
final states when they interact in targets.  

As there is no obvious quantum number distinguishing the $\nu_e$ and $\bar{\nu}_e$,
would would then be tempted to introduce one, the lepton number $l_e$, 
requiring it to be additively conserved:
\[ \begin{array}{cc} \mathrm{lepton} &~~ l_e \\
e^- & +1 \\
e^+ & -1 \\
\nu_e & +1 \\
\bar{\nu}_e & -1 \\
\end{array} ~~~~\Leftrightarrow ~~~~\sum_{in} l_e = \sum_{out} l_e .\]
This would account for the results of the ``experiments" illustrated in Fig.~\ref{fig:nue}.
Historically this issue was connected with the early development of the Cl detector
famous in solar neutrinos.  After Pontecorvo suggested using Cl, 
Alvarez investigated
various background issues, as he was considering testing $\nu$/$\bar{\nu}$
identity, using reactor $\bar{\nu}_e$s.  Later Davis placed a Cl detector prototype near the
Savannah River reactor to search for
\[ {}^{37} \mathrm{Cl} +\bar{\nu}_e \rightarrow {}^{37}\mathrm{Ar} + e^- \]
but found no Ar, indicating that the $\nu_e$ and $\bar{\nu}_e$ are distinct at the
level of $\sim$ 5\%.

\begin{figure}
\begin{center}
\includegraphics[width=11cm]{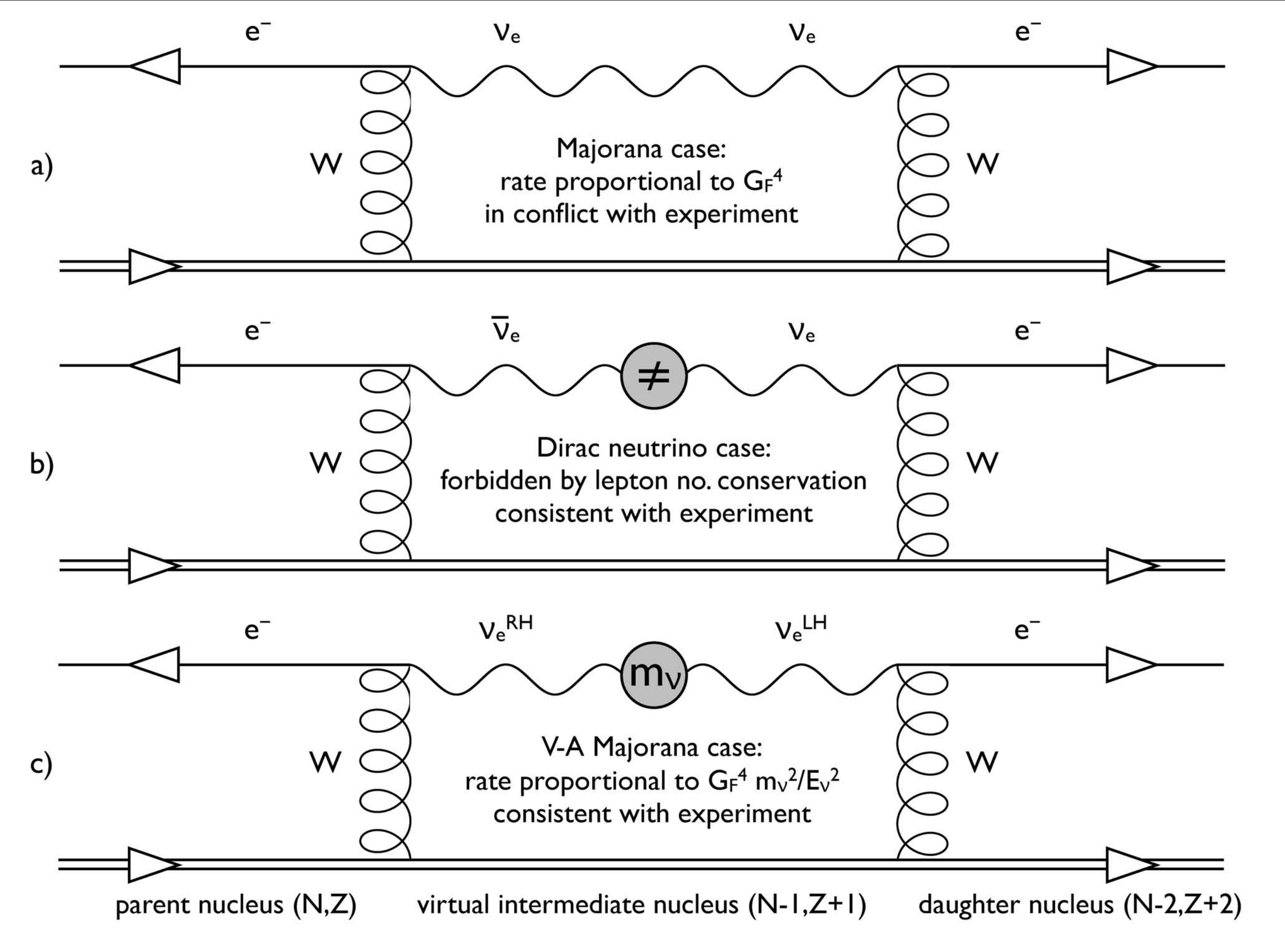}
\end{center}
\caption{ Panel a) neutrinoless double beta decay in the case of a Majorana
neutrino appears naively to be in conflict with experiment, as it produces a
substantial decay rate.  In the Dirac case (Panel b)) the decay is strictly
forbidden by lepton number conservation: the $\bar{\nu}_e$ produced in the first decay
is the wrong neutrino to complete the second step.  A Dirac neutrino is
thus allowed by experiment, as it is consistent with the absence of neutrinoless
double beta decay, a process so far not seen in nature.
Panel c) shows the Majorana case, including the effects
of neutrino handedness.  A Majorana neutrino is allowed by experiment,
provided the mass is sufficiently small (so the handedness is sufficiently exact).
This requires the Majorana neutrino mass to be $\lsim$ 1 eV, a result compatible
with other tests of mass, such as neutrino oscillations.}
\label{fig:doublebeta}
\end{figure}

There is an elegant way to do the Savannah River experiment at
the nuclear level, with neutrinoless $\beta \beta$ decay \cite{bb}.
In this process a nucleus (N,Z) decays through a second-order
weak interaction
\[ (N,Z) \rightarrow (N-2,Z+2) + e^- + e^- \]
with the emission of two electrons.  If the $\nu_e$ and $\bar{\nu}_e$ are identical $-$ if the
neutrino is a Majorana particle $-$
then the neutrino emitted in one neutron $\beta$
decay can be reabsorbed on a second neutron, as shown in the upper panel
of Fig.~\ref{fig:doublebeta}.   But we do not see this process in nature.
Thus it seems once again that 
the $\nu_e$ and $\bar{\nu}_e$ must be distinct, carrying opposite lepton charges.
As the middle panel illustrates, 
neutrinoless double $\beta$ decay would then be strictly 
forbidden: the final state of two electrons
($l_e=2$) cannot result from the decay of an initial state with $l_e=0$.

But there is a hidden assumption in the above arguments, one that starts with the
axial current introduced by Gamow and Teller.  We can remove all references
to lepton number $-$ so that $|\bar{\nu}_e \rangle = |\nu_e \rangle$ $-$ but suppress
double $\beta$ decay with a different label, the handedness of the neutrino.  As
the Goldhaber, Grodzins, and Sunyar experiment would show in 1957, because 
the Gamow and Teller spin interaction does appear with Fermi's vector interaction
in the combination V-A, the (anti)neutrino emitted in $\beta^-$ decay is righted-handed,
while the neutrino that produces an $e^-$ when scattering in a target is left-handed.
Thus, in the third panel of Fig.~\ref{fig:doublebeta}, we find another mechanism
for suppressing double $\beta$ decay, without the assumption of a conserved
lepton number: the neutrino produced in the first beta decay has the wrong handedness
to generate a neutrino by scattering off a second nucleon.

These assignments do not forbid double $\beta$ decay $-$ we eliminated lepton
number and its absolute conservation $-$ but only suppress it: the handedness of
the neutrino is not exact because of its mass.   The amplitude in 
panel c) is proportional to the handedness admixture
\[ {m_\nu \over E_\nu}  \sim 3 \times 10^{-8}~{m_\nu \over 1~ \mathrm{eV}},\]
where we have estimated the energy of the exchanged neutrino to be
$E_\nu \sim \mathrm{R}_\mathrm{nucleus}^{-1} \sim 6 f^{-1}$ for a typical heavy
nucleus. The same arguments apply to the Savannah River experiment and other
tests that naively suggest the $\nu$ must have a distinct antiparticle.

\begin{figure}
\begin{center}
\includegraphics[width=11cm]{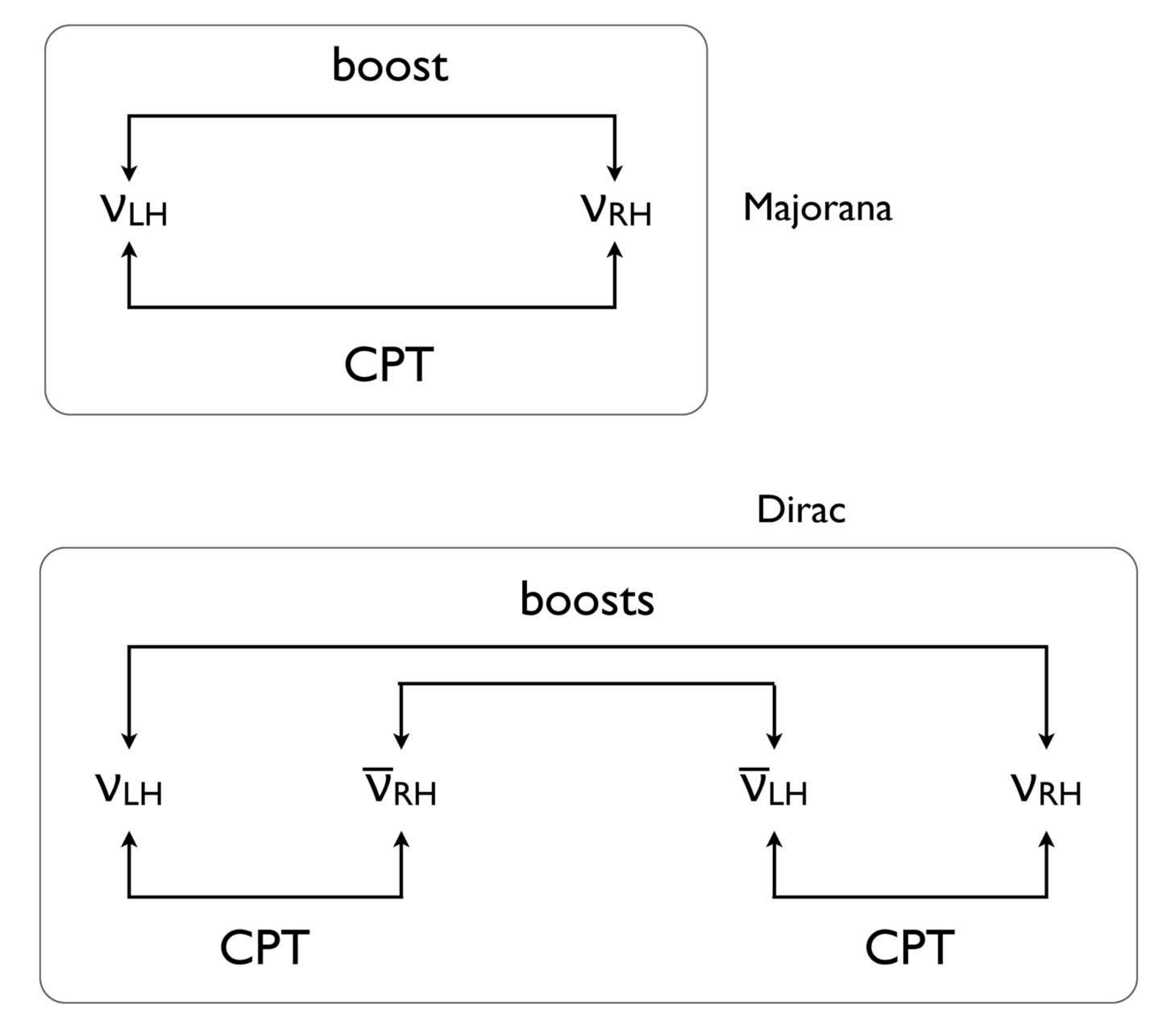}
\end{center}
\caption{ The two-component (Majorana) and four-component (Dirac) neutrino
schemes, showing the relationships between the components under Lorentz
boosts and particle-antiparticle conjugation.}
\label{fig:CPT}
\end{figure}

Thus this discussion shows that there are two descriptions of massive neutrinos,
and {\it both} are consistent with nature, for light neutrinos.  These are illustrated in
Fig.~\ref{fig:CPT}.  The first is the two-component Majorana case: the neutrino
is its own antiparticle, so that under CPT the $|\nu_{LH} \rangle \rightarrow |\nu_{RH} \rangle$.
One can also consider the constraints of Lorentz invariance.   If one boosts to a frame
moving faster than a massive neutrino, momentum is reversed but not spin. 
The handedness is then flipped: Lorentz invariance requires a right-handed counterpart
to a left-handed state, and the reverse.   Thus a two-component neutrino can satisfy
the constraints of both particle-antiparticle conjugation and Lorentz invariance. 

Alternatively, one can introduce a lepton number to distinguish the $\nu$ and $\bar{\nu}$.
Lepton number reverses under particle-antiparticle conjugation, but not under
Lorentz boosts.  Thus four neutrino components are required to satisfy the 
constraints of particle-antiparticle conjugation and Lorentz invariance.  This is the
Dirac case of Fig.~\ref{fig:CPT}.

Our minimal standard model of particle physics does not allow neutrino mass: Dirac
masses are absent because there is no right-handed neutrino field, and Majorana masses
are absent because they require interactions that are poorly behaved at high
energies.  But neither of these standard-model restrictions is likely to hold in the
more general models that will someday replace the standard model; nor is the second
concern of any importance if one adopts a modern view that the standard model is an
effective theory that should be restricted to lower energies.

We now know a new, extended standard model is needed, because neutrinos are massive.
The Super-Kamiokande and SNO experiments have demonstrated that
atmospheric and solar neutrinos oscillate.  While oscillation experiments constrain
differences in the squares of neutrino masses, those differences tell us that at
least one neutrino has a mass $ m \gsim$ 0.04 eV.  We also have the $\nu$ mass bounds
\[ \sum_i m_i \lsim 6.6~ \mathrm{eV}~~\mathrm{(laboratory)}~~~~~~\sum_i m_i \lsim 0.7 ~\mathrm{eV}~~\mathrm{(cosmology)} \]
The former comes from combining tritium $\beta$ decay mass limits with mass differences
$\delta m^2 = m_i^2-m_j^2$ known from oscillation experiments, while the later is derived
from cosmological arguments, that massive neutrinos inhibit the growth of structure
on large scales.

This chain of argument leads to an interesting observation: while we must look to extended
models for an explanation of neutrino mass, it is at first not clear how a more unified
theory can account for neutrino masses, since they are so much smaller than the masses
of other standard model particles.  A group theory factor is not going to generate the factor
of 10$^{-6}$ or more to explain the ratio $m_{\nu_e}/m_e$.  But a elegant resolution of this
puzzle is provided by the special freedom available with neutrinos, the
possibility of both Dirac and Majorana mass terms \cite{seesaw}.  Neutrinos and other
standard-model fermions can share the same Dirac mass scale $m_D$.  But the seesaw
mechanism, in addition, postulates a heavy right-handed Majorana neutrino mass
$m_R$, yielding a mass matrix
\[ \left( \begin{array}{cc} 0 & m_D \\ m_D & m_R \end{array} \right) \begin{array}{c} \Rightarrow \\
~~~~\small{\mathrm{diagonalize}}~~~~ \end{array} m_\nu^{\mathrm{light}} \sim m_D \left( {m_D \over M_R} \right) \]
Thus a natural ``small parameter" $m_D/m_R$ emerges that explains why neutrinos are
so much lighter than other fermions.  The scale $m_R$ represents new physics.  
If one takes
\[ \left. \begin{array}{l} m_\nu \sim \sqrt{\delta m^2_{\mathrm{atmos.}}} \sim 0.05~\mathrm{eV} \\
m_D \sim m_{\mathrm{top}} \sim 180~ \mathrm{GeV} \end{array} \right\} ~~\Rightarrow~~
m_R \sim 0.3 \times 10^{15} ~\mathrm{GeV}, \]
one finds an $m_R$ that is very close to the energy scale where supersymmetric Grand
Unified theories  predict that the strong, weak, and electromagnetic forces unify.  Thus it
is quite possible that tiny neutrino masses are giving us our first glimpse of physics that
is otherwise hidden by an enormous energy gap.

\section{The Goldhaber-Teller Model}
\begin{figure}
\begin{center}
\includegraphics[width=11cm]{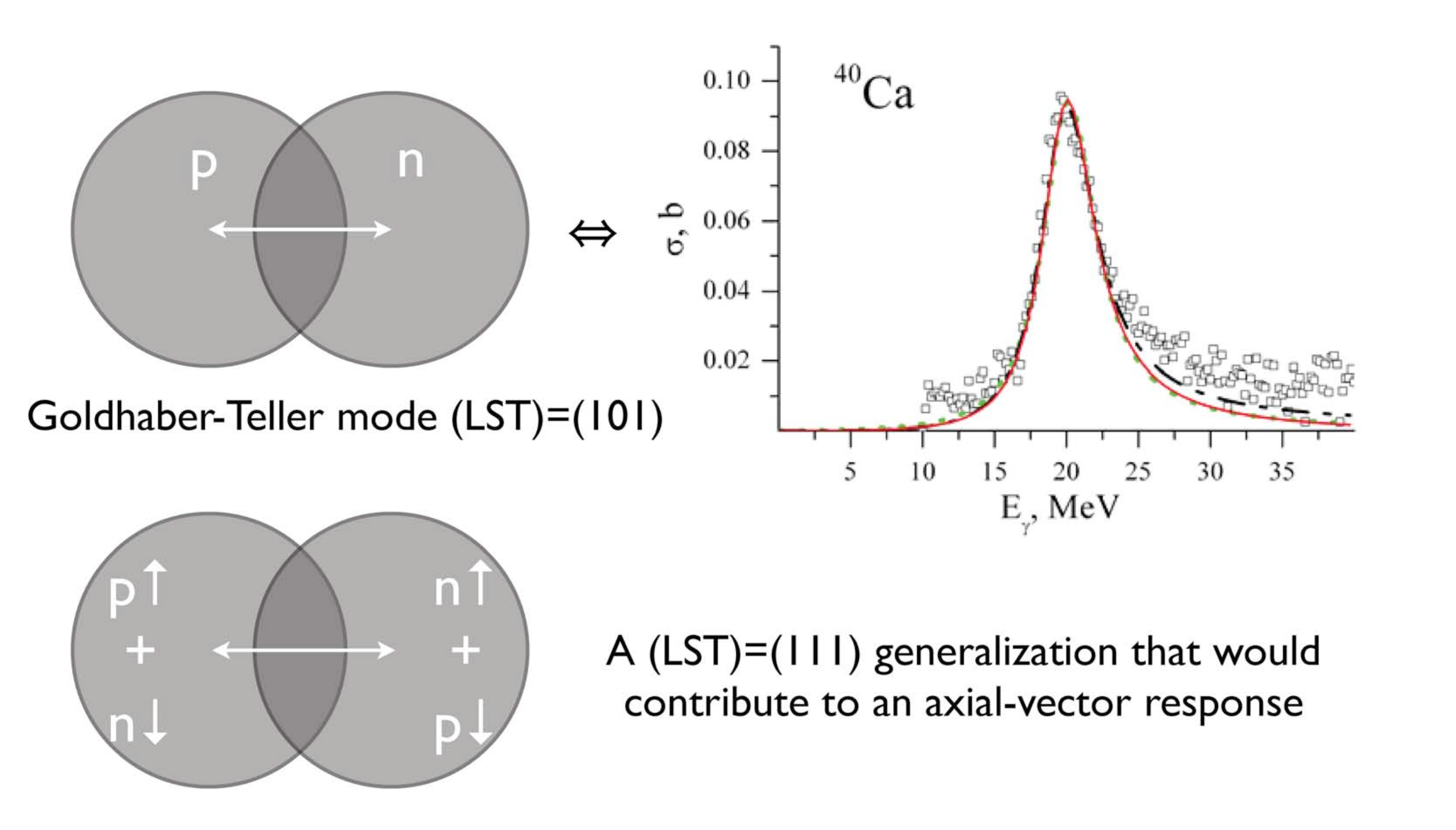}
\end{center}
\caption{ Schematic illustrations of the original Goldhaber-Teller E1 mode, saturating the
photoabsorption sum rule, and of an S=1 generalization, one of the giant dipole modes
that would be excited by the axial-vector current in inelastic neutrino scattering.}
\label{fig:Goldhaber}
\end{figure}

In 1946 Edward Teller left Los Alamos to join Enrico Fermi and Maria Mayer at the University
of Chicago.  In 1948 Maurice Goldhaber, then at the University of Illinois,
approached Teller about the broad photoabsorption resonances he had observed in
$(\gamma$,n) nuclear reactions -- the nuclear response was similar to related phenomena in crystal lattices that Teller had considered years earlier.   Out of these discussions emerged
a simple collective model of the giant dipole resonance in which the neutrons oscillate
against the protons, with the nuclear symmetry energy generating a linear restoring force.
This single harmonic
mode could be constructed to satisfy the energy-weighted $E1$ sum rule
\[ {2M \over \hbar^2} \sum_f (E_f-E_{g.s.}) \langle f | \sum_{i=1}^{A} z(i) {\tau_3 \over 2} |g.s. \rangle^2 = {A \over 4}. \]
Figure~\ref{fig:Goldhaber} illustrates the mode.

This celebrated paper \cite{goldteller} also has a connection with neutrinos and with the Gamow-Teller
suggestion of a semi-leptonic weak interactions mediated by a combination of vector and
axial-vector currents.  Once the model is constructed to saturate the $E1$ sum rule,
it can be generalized algebraically to account for the full 
15-dimensional super-multiplet of dipole resonances,
if it is assumed nuclear forces are independent of spin and isospin.   This
yields a set of SU(4) generalizations of the $E1$ mode which saturate the (L=1 S T) responses
for the allowed choices of spin S and isospin T.

One can envision this pictorially as a generalization of the proton/neutron fluid oscillations of
the original Goldhaber-Teller model to fluids that also carry spin.  For example, the axial responses
needed to describe neutral current neutrino scattering involve the substitution
\[ z(i) \tau_3(i) ~~\Rightarrow~~\left[ \vec{r}(i) \otimes \vec{\sigma}(i) \right]_{J=0,1,2} \left( \begin{array}{c} \tau_3(i) \\ 1 \end{array} \right) \]
An example of such a spin-dependent mode is given in Fig.~\ref{fig:Goldhaber}:  a fluid
consisting of spin-up protons and spin-down neutrons oscillates against one with spin-up
neutrons and spin-down protons.  Thus this is an L=1 S=1 T=1 mode.

Despite its great simplicity, this generalized Goldhaber-Teller model captures enough of the
basic physics of first-forbidden nuclear responses to be a useful tool for calculations.
One ``homework" problem \cite{woosley} that can be solved with this model is ``How did the galaxy make the
fluorine found in toothpaste?"  That data needed for the solution include:
\begin{itemize}
\item Fluorine has a single stable isotope, ${}^{19}$F, and is relatively rare.   Its galactic
abundance relative to neon is 1/3100.
\item Galactic core-collapse supernovae occur about once every 30 years.
\item Such supernovae are thought to be the primary source of Ne, which is synthesized from
``burning" oxygen in the supernova progenitor star during the long period of hydrostatic
evolution, then ejected with the rest of the mantle during the explosion.
\item Each supernova releases about $4 \times 10^{57}$ muon and tauon $\nu$s.
\item As these $\nu$s have an average energy of $\sim$ 25 MeV, most can excited 
giant resonances when scattering off Ne.
\item As the typical radius of the Ne shell is about 20,000 km, the total fluence of $\nu$s
through this shell during the few seconds of the explosion is $\sim 10^{38}$/cm$^2$.
\end{itemize}

Although the average neutrino energy $\omega \sim$ 25 MeV,
cross section kinematics yields makes neutrinos on the high-energy tail of the Boltzmann
distribution much more important.   Furthermore, neutrinos scattering  backward
can transfer a three-momentum $|\vec{q}| \sim 2 \omega$.  Consequently the
inelastic scattering is dominated by the first-forbidden responses, the giant resonances.
While the original Goldhaber-Teller model was designed to saturate the sum rule for the
$(\gamma,n)$ nuclear response, its SU(4) generalization addresses the full set of
giant resonances important to both axial and vector currents.  That is, it provides a
reasonable model for
\[ {}^{20}Ne(Z_0,n)^{19}Ne \rightarrow {}^{19}F ~~~\mathrm{or}~~~{}^{20}Ne(Z_0,p)^{19}F. \]
where $Z_0$, the boson mediating neutrino-nucleus scattering, is the analog of the $\gamma$
in Goldhaber's photoexcitation experiments.  Summing over the spin and isospin modes yields
\[ \sigma \sim 3 \times 10^{-41}~ \mathrm{cm}^2 \Rightarrow~\mathrm{1/300~of~Ne~shell~converted~to~F} \]

The direct synthesis of new elements by supernova neutrinos is called the neutrino process.
When calculations only slightly more sophisticated than the above are combined with
nuclear reaction networks that include the effects of the processing of the neutrons
and protons coproduced with the ${}^{19}$Ne and ${}^{19}$F, as well as effects of 
shock-wave heating of the Ne shell, one finds that the correct ratio of ${}^{19}$F/${}^{20}$Ne
is obtained for a heavy-flavor neutrino temperature of $\sim$ 6 MeV.  This is very similar
to the temperatures supernova modelers obtain from detailed hydrodynamic simulations.

\section{Summary}
Edward Teller's career in nuclear physics spanned 70 years and encompassed 
many subjects not addressed here.  However, the two papers chosen for discussion
are notable for their physics insight and their historical setting.

Gamow and Teller introduced a spin-dependent $\beta$ decay amplitude that
anticipated the axial-vector current of the standard model and the 1957 discoveries
that established the V-A form of the weak interaction.   The resulting handedness of
the neutrino provides an explanation for suppressed double $\beta$  decay rates,
thereby keeping open the possibility of Majorana neutrino masses.   Because
neutrinos can have both Majorana and Dirac mass terms, a natural explanation
arises for the difference between the neutrino mass scale and that of other fermions --
but one not yet confirmed experimentally.

Goldhaber and Teller introduced a simple nuclear model capable of saturating the
$E1$ photoabsorption sum rule.  This model has a natural generalization for weak 
charged- and neutral-current responses, if the nuclear Hamiltonian is assumed be
independent of spin and isospin.  While the Gamow-Teller paper described for the
first time the full set of operators responsible for allowed transitions, and Goldhaber-Teller
model and its SU(4) generalization provided a similar description for the first-forbidden
responses.

These two papers together provide a framework for tackling astrophysics problems that 
range from the pp-chain and Big-Bang nucleosynthesis to the interactions of 
neutrinos in the mantle of a supernova.

\section{Acknowledgements}
This work was supported in part by the U.S. Department of Energy, Office of Nuclear Physics,
under grant \#DE-FG02-00ER41132.

\end{document}